\documentclass[11pt]{article}

\usepackage[superscript,sort]{cite}
 
\usepackage{ifthen,ifpdf}
\ifpdf
	\usepackage[pdftex]{hyperref}
	\hypersetup{colorlinks=true,linkcolor=black,citecolor=black,urlcolor=blue,backref=page,bookmarks=true,breaklinks=true,plainpages=false}
	\usepackage[pdftex]{graphicx}
	\usepackage[usenames,pdftex]{color}
\else
	\usepackage[colorlinks=true,breaklinks=true,plainpages=false]{hyperref}
	\usepackage{graphicx}
	\usepackage[usenames]{color}
\fi

\usepackage{amsthm,amscd,amsxtra,amsfonts,amsmath,amssymb,multirow}
\usepackage{wrapfig}

\usepackage{helvet}

\usepackage[letterpaper,top=1.0in,bottom=1.0in,left=1.0in,right=1.0in]{geometry}

\definecolor{MyPurple}{rgb}{1,0,1}

\begin{document}
	
\title{Differential geometry-based solvation and electrolyte transport models for biomolecular modeling: a review.}

\author{Guo Wei Wei\footnote{Department of Mathematics, Department of Biochemistry and Molecular Biology, Michigan State University, MI 48824, USA. wei@math.msu.edu}
\and 
Nathan A.~Baker\footnote{Computational and Statistical Analytics Division, Pacific Northwest National Laboratory, Richland, WA 99352, USA. nathan.baker@pnnl.gov}}

\date{\today}
\maketitle

\begin{abstract}
This chapter reviews the differential geometry-based solvation and electrolyte transport for biomolecular solvation that have been developed over the past decade. A key component of these methods is the differential geometry of surfaces theory, as applied to the solvent-solute boundary. In these approaches, the solvent-solute boundary is determined by a variational principle that determines the major physical observables of interest, for example, biomolecular surface area, enclosed volume, electrostatic potential, ion density, electron density, etc.
Recently, differential geometry theory has been used to define the surfaces that separate the microscopic (solute) domains for biomolecules from the macroscopic (solvent) domains.
In these approaches, the microscopic domains are modeled with atomistic or quantum mechanical descriptions, while continuum mechanics models (including fluid mechanics, elastic mechanics, and continuum electrostatics) are applied to the macroscopic domains.
This multiphysics description is integrated through an energy functional formalism and the resulting Euler-Lagrange equation is employed to derive a variety of governing partial differential equations for different solvation and transport processes; e.g., the Laplace-Beltrami equation for the solvent-solute interface, Poisson or Poisson-Boltzmann equations for electrostatic potentials, the Nernst-Planck equation for ion densities, and the Kohn-Sham equation for solute electron density.
Extensive validation of these models has been carried out over hundreds of molecules, including proteins and ion channels, and the experimental data have been compared in terms of solvation energies, voltage-current curves, and density distributions. We also propose a new quantum model for electrolyte transport. 

\end{abstract}

 
\pagestyle{empty}
 
\newpage
{\setcounter{tocdepth}{3} \tableofcontents \listoffigures}

\newpage

\section{Background}
Solvation is an elementary process in nature and is particularly essential to biology.
Physically, the solvation process can be described by a variety of interactions, such as electrostatic, dipolar, induced dipolar, and van der Waals, between the solvent and solute.
Due to the ubiquitous nature of electrostatics and the aqueous environment common to most biomolecular systems, molecular solvation and electrostatics analysis is significantly important to research in chemistry, biophysics, and medicine. Such analyses can be classified into two general types: 1) quantitative analysis for thermodynamic or kinetic observables and 2) qualitative analysis for general characteristics of biomolecular solvation. 

In general, implicit solvent models describe the solvent as a dielectric continuum, while the solute molecule is modeled with an atomistic description \cite{Baker:2005a,Davis:1990a,Sharp:1990a,Honig:1995a,Roux:1999,Jinnouchi:2008}.
There are many two-scale implicit solvent models available for electrostatic analysis of solvation, including generalized Born (GB) \cite{Dominy:1999,Bashford:2000,Tsui:2000,Onufriev:2002,Gallicchio:2002,Zhu:2005,Koehl:2006,Tjong:2007b,Mongan:2007,Chen:2008,Grant:2007}, polarizable continuum    \cite{Chiba:2008,Tomasi:2005,Improta:2006,Takano:2005,Cances:1997,Barone:1997,Cossi:1996} and Poisson-Boltzmann (PB) models \cite{Lamm:2003,Fogolari:2002,Sharp:1990a,Davis:1990a,Zhou:2008b,Baker:2005}.
GB methods are fast heuristic models for approximating polar solvation energies.
PB methods can be formally derived from basic statistical mechanics theories for electrolyte solutions \cite{Beglov:1996,Netz:2000a,Holm:2001} and therefore offer the promise of  robust models for computing the polar solvation energy \cite{David:2000,Onufriev:2000,Bashford:2000}.
In many solvation analyses, the total solvation energy is decomposed into polar and nopolar contributions.
Although there are many ways to perform this decomposition, many approaches model the nonpolar energy contributions in two stages:  the work of displacing solvent when adding a hard-sphere solute to solution and the dispersive nonpolar interactions between the solute atoms and surrounding solvent.

One of the primary quantitative applications of implicit solvent methods in computational biology and chemistry research has involved the calculation of thermodynamic properties. 
Implicit solvent methods offer the advantage of ``pre-equilibrating'' the solvent and mobile ions, thus effectively pre-computing the solvent contribution to the configuration integral or partition function for a system \cite{Roux:1999}.
Such pre-equilibration is particularly evident in molecular mechanics/Poisson-Boltzmann surface area (MM/PBSA) models \cite{Weinzinger:2005,Swanson:2004,Page:2006c,Tan:2006c,Massova:1999} that combine implicit solvent approaches with molecular mechanics models to evaluate biomolecule-ligand binding free energies from an ensemble of biomolecular structures.
The calculation and assignment of protein titration states is another important application of implicit solvent methodology \cite{Bashford:1990,Antosiewicz:1996,Li:1996,Nielsen:2001,MacDermaid:2007,Tang:2007,Nielsen:1999,Yang:1993a,Nielsen:2001,Georgescu:2002,Matousek:2007,PROPKA,Li:2004e}. 
Such methods have been used to interpret experimental titration curves, decompose residue contributions to protein-protein and protein-ligand binding energetics, examine structural/functional consequences of RNA nucleotide protonation, as well as several other applications.
Another application area for implicit solvent methods is in the evaluation of biomolecular dynamics, where implicit solvent models generally are used to provide solvation forces for molecular Langevin dynamics \cite{Tan:2006d,Prabhu:2008,Prabhu:2004,AMBER-PB,Lu:2003, Geng:2011}, Brownian dynamics \cite{UHBD,Gabdoulline:1998,Elcock:1999,Sept:1999}, or continuum diffusion \cite{Cheng:2007,Cheng:2007a,Zhang:2005,Song:2004b,Song:2004} simulations.
A major qualitative use of implicit solvent methods in experimental work is the visualization and qualitative analysis of electrostatic potentials on and around biomolecular surfaces \cite{Warwicker:1982,GRASP2,Baker:2003,Baker:2004}. 
Visualization of electrostatic potentials was popularized by the availability of software, such as Grasp \cite{GRASP2}, and is now a standard procedure for analyzing biomolecular structures with thousands of examples available in the literature, including ligand-receptor binding and drug design, protein-nucleic acid complexes, protein-protein interactions, macromolecular assembly, and enzymatic mechanism analysis, among others.
More complete descriptions of the solvation process, solvation models, and various applications of solvation methods also can be found in the literature \cite{ZhanChen:2010a,ZhanChen:2010b,Ren:2013}.
Typically, solvation models are tested against experimental data for solvation free energies, titration and redox behaviors, or spectroscopic measures of local electric fields.
However, solvation models can also provide insight into molecular properties which cannot be directly measured experimentally, including solute surface area and enclosed volume, electrostatic potential, and nonpolar solvation behavior.
The properties derived from solvation models are used in a variety of applications, including pH and p$K_a$ estimation, titration analysis, stability analysis, visualization, docking, and drug and protein design.
In addition, sophisticated models for non-equilibrium processes, such as Brownian dynamics, molecular dynamics, kinetic models, and multiscale models, may have a solvation model as a basic component \cite{Wei:2009,Wei:2012,Wei:2013}.  
 
\section{Differential geometry-based solvation models} \label{solvation-model}
Most implicit solvent models require a definition of the solvent density and/or dielectric coefficient profile around the solute molecule.
Often, these definitions take the form of analytic functions \cite{Zap,Grant:1995,Grant:2007} or discrete boundary surfaces dividing the solute-solvent regions of the problem domain.
The van der Waals surface, solvent accessible surface \cite{Lee:1971}, and molecular surface (MS) \cite{Richards:1977} are typically used for this purpose and have found many successful applications in biomolecular modeling \cite{Spolar:1989,Livingstone:1991,Crowley:2005,Kuhn:1992,Bergstrom:2003,Dragan:2004,Jackson:1995,Licata:1997}.
Physical properties calculated from implicit solvent models are very sensitive to the definition of the dielectric profile \cite{Dong:2003,Dong:2006,Nina:1999,Swanson:2005a}; however, many of these popular profile definitions are \textit{ad hoc} divisions of the solute and solvent regions of the problem domain based on assumptions about molecular geometry rather than minimization of solute-solvent energetic interactions.

Geometric analysis, which combines differential geometry (DG) and differential equations, has had a tremendous impact in signal and image processing, data analysis, surface construction \cite{Feng:2004a,Gomes:2001,Mikula:2004,Osher:2001,Sarti:2002,Sbert:2003,Sethian:2001,Sochen:1998}, and surface smoothing \cite{Zhang:2006c}.
Geometric partial differential equations (PDEs) \cite{Willmore:1997}, particularly mean curvature flows, are popular tools in applied mathematics.
Computational techniques using the level set theory were devised by Osher and Sethian \cite{SOsher:1988,Rudin:1992,Sethian:2001} and have been further developed and applied by many others \cite{Cecil:2005,Chopp:1993,Smereka:2003}.
An alternative approach is to minimize the mean curvature or energy functional of the hypersurface function in the framework of the Mumford-Shah variational functional \cite{Mumford:1989}, and the Euler-Lagrange formulation of surface variation developed by Chan and coworkers, and others \cite{Blomgren:1998,Carstensen:1997,Li:1996a,Osher:1990,Rudin:1992,Sapiro:1996}.
Wei introduced some of the first high-order geometric PDEs for image analysis \cite{Wei:1999} and, with coworker Jia, also presented the first geometric PDE-based high-pass filters by coupling two nonlinear PDEs \cite{Wei:2002a}.
Recently, this approach has been generalized to a more general formalism, the PDE transform, for image and surface analysis \cite{YWang:2011b,YWang:2011c,YWang:2011d}, including biomolecular surface generation \cite{QZheng:2012}.  

Geometric PDEs and DG theories of surfaces provide a natural and simple description for a solvent-solute interface.
In 2005, Wei and his collaborators, including Michael Feig, pioneered the use of curvature-controlled PDEs for molecular surface construction   and solvation analysis \cite{Wei:2005}.
In 2006, based on DG, Wei and coworkers introduced  the first variational solvent-solute interface: the minimal molecular surface (MMS),   for molecular surface representation  \cite{Bates:2006,Bates:2006f,Bates:2008}.
With a constant surface tension, the minimization of surface free energy is equivalent to the minimization of surface area, which can be implemented via the mean curvature flow, or the Laplace-Beltrami flow, and gives rise to the MMS. 
The MMS approach has been used to calculate both solvation energies and electrostatics \cite{ZhanChen:2012, Bates:2008}.
Potential-driven geometric flows, which admit non-curvature-driven terms, have also been proposed for biomolecular surface construction \cite{Bates:2009}.
While our approaches were employed by many others \cite{Cheng:2007e,Yu:2008g,SZhao:2011a,SZhao:2014a} for molecular surface analysis, our curvature-controlled PDEs and the geometric flow-based MMS model proposed in 2005 \cite{Wei:2005,Bates:2006,Bates:2008,Bates:2009} are, to our knowledge, the first of their kind for biomolecular surface and electrostatics/solvation modeling.  

\begin{figure}[p]
	\centering
	\includegraphics[keepaspectratio,width=0.8\textwidth]{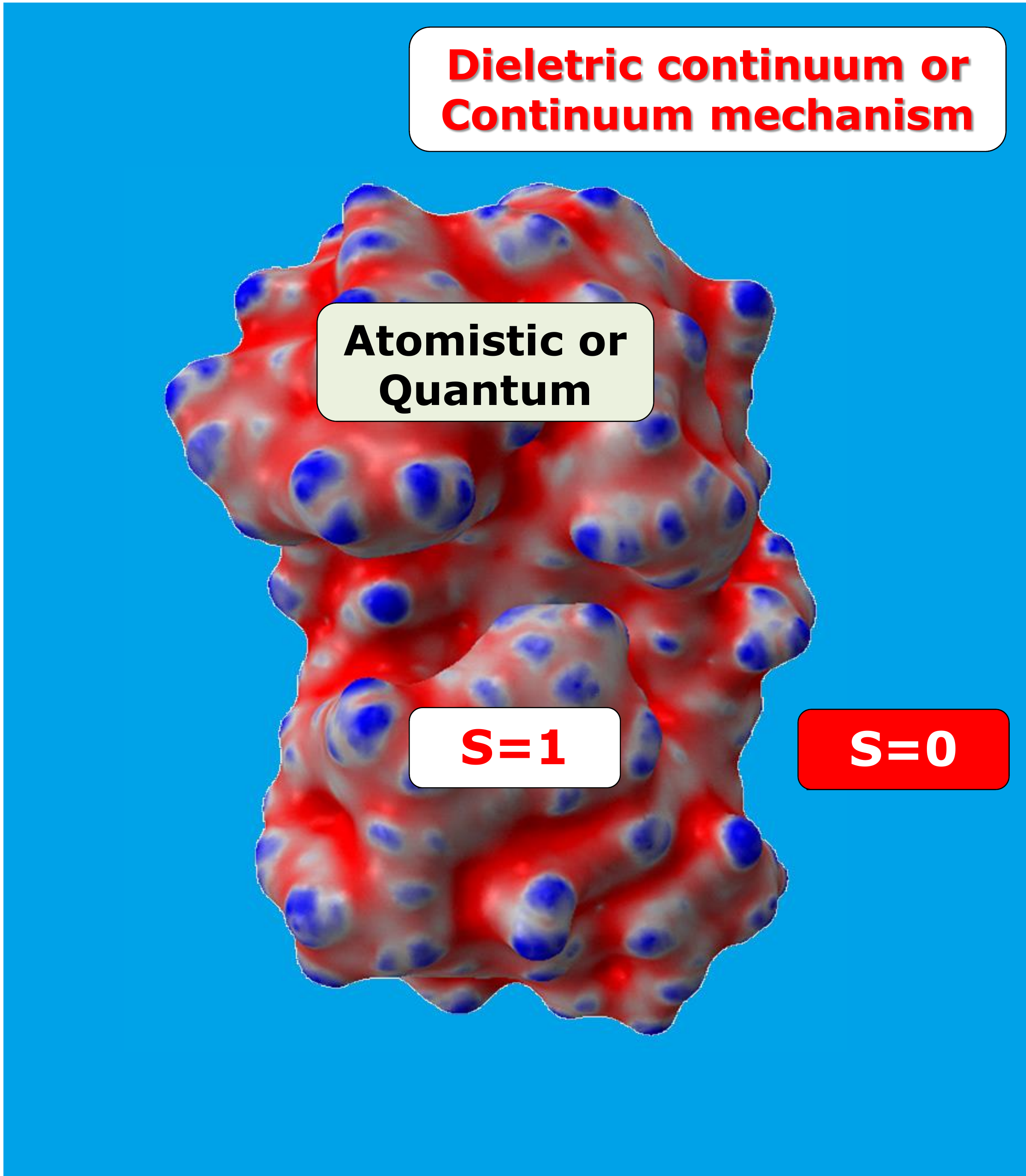}
	\caption{An illustration of differential geometric based solvation models. The minimum curvature is mapped on the Laplace-Beltrami surface of protein penicillopepsin (PDB ID 2web).}
	\label{fig:solvation}
\end{figure}

Our DG theory of the solvent-solute interface can be extend into a full solvation model by incorporating a variational formulation of the PB theory \cite{Sharp:1990, Gilson:1993} as well as a model of nonpolar solute-solvent interactions \cite{ZhanChen:2012} following a similar approach by Dzubiella, Swanson, and McCammon \cite{Dzubiella:2006}.
We have implemented our DG-based solvation models in the Eulerian formulation, where the solute boundary is embedded in the three-dimensional (3D) Euclidean space so evaluation of the electrostatic potential can be carried out directly \cite{ZhanChen:2010a}.
We have also implemented our DG-based solvation models in the Lagrangian formulation \cite{ZhanChen:2010b} (see Fig.~\ref{fig:solvation}) wherein the solvent-solute interface is extracted as a sharp surface and subsequently used in solving the PB equation for the electrostatic potential.
To account for solute response to solvent polarization, we recently introduced a quantum mechanical (QM) treatment of solute charges to our DG-based solvation models using density functional theory (DFT) \cite{ZhanChen:2011a}.
Most recently,   Wei and coworkers have taken a different treatment of non-electrostatic interactions between the solvent  and solute in the DG based solvation models so that the resulting total energy functional and PB equations are consistent with more detailed descriptions of solvent densities at equilibrium \cite{Wei:2012,Wei:2013}.  
This multiscale approach self-consistently computes the solute charge density distribution which simultaneously minimizes both the DFT energy as well as the solvation energy contributions.
The resulting model significantly extends the applicability of our solvation model to a broad class of molecules without the need for force-field parametrized charge terms.
The resulting differential geometry implicit solvent model has been tested extensively and shows excellent performance when compared with experimental and explicit solvent reference datasets \cite{ZhanChen:2010a, ZhanChen:2010b, ZhanChen:2011a, ZhanChen:2012, DuanChen:2012a, DuanChen:2012b, Wei:2012, Daily:2013, Thomas:2013}.

As mentioned above, a parallel line of research has been carried out by Dzubiella, Hansen, McCammon, and Li.
Early work by Dzubiella and Hansen demonstrated the importance of the self-consistent treatment of polar and nonpolar interactions in solvation models \cite{DzubiellaHansen:2003, DzubiellaHansen:2004}.
These observations were then incorporated into a self-consistent variational framework for polar and non-polar solvation behavior by Dzubiella, Swanson, and McCammon \cite{Dzubiella:2006, Dzubiella:2006a} which shared many common elements with our earlier geometric flow approach but included an additional term to represent nonpolar energetic contributions from surface curvature.  Li and co-workers then developed several mathematical methods for this variational framework based on level-set methods and related approaches \cite{Che:2008, Cheng2:2009, Li2013Variational} which they demonstrated and tested on a variety of systems \cite{Setny:2009, Cheng2009Interfaces, Zhou2013Variational}. Unlike our Eulerian representation \cite{ZhanChen:2010a}, level-set methods typically give rise to models with sharp solvent-solute interfaces.

\begin{figure}[p]
	\centering
	\includegraphics[keepaspectratio,width=0.8\textwidth]{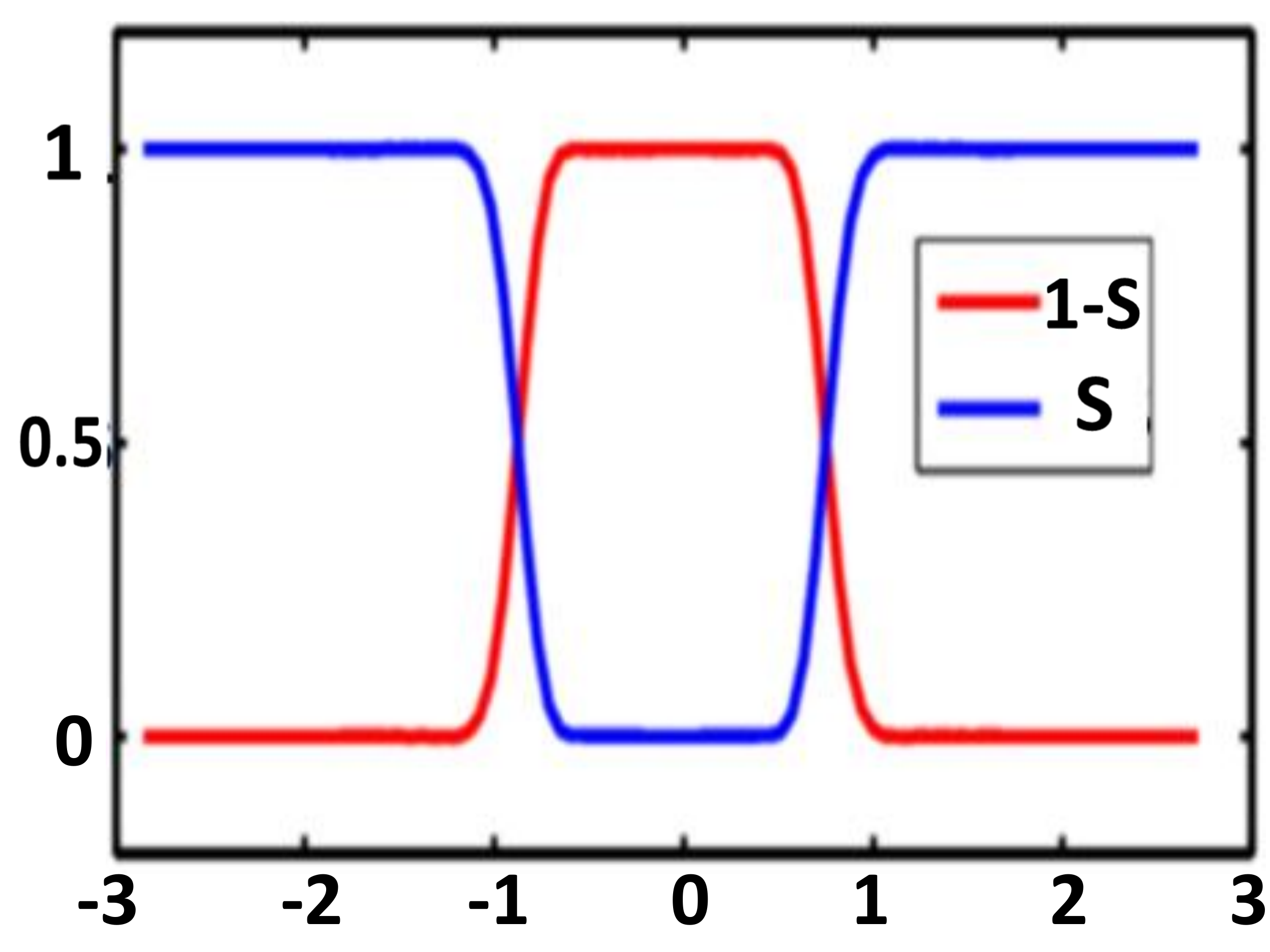}
	\caption{An illustration of the one-dimensional projection of the profiles of $S$ and $1-S$ functions along the $x$-axis. }
	\label{fig:S-profile}
\end{figure}

An immediate consequence of our models is that the surfaces generated are free of troublesome geometric singularities that commonly occur in conventional solvent-accessible and solvent-excluded surfaces \cite{Connolly85,Sanner:1996} and impact computational stability of methods (see Fig.~\ref{fig:S-profile} for a smooth surface profile).
Addition, without using {\it ad hoc} molecular surfaces, both our solvation models and the models of Dzubiella {\it et al.}~significantly reduce the number of free parameters that users must ``fit'' or adjust in applications to real-world systems \cite{Thomas:2013}.
Our recent work shows that physical parameters; i.e., pressure and surface tension, obtained from experimental data can be directly employed in our DG-based solvation models to achieve an accurate prediction of solvation energy \cite{Daily:2013}.

In this chapter, we review a number of DG-based models. Initially, we discuss solvation models, i.e, nonpolar and polar solvation models at equilibrium. To improve the accuracy and make our models robust, quantum mechanics is applied to the solute's electron structure. As an important extension, we also consider DG-based models for the dynamical processes at non-equilibrium settings, including applied external electrical field gradients and inhomogeneous solvent concentration across membrane proteins. 

\subsection{Nonpolar solvation model}\label{Nonpolar}
As discussed above, solvation free energy is typically divided into two contributions: polar and nonpolar components.
In one popular description, polar portion refers to electrostatic contributions while the nonpolar component includes all other effects.
Scaled particle theory (SPT) is often used to describe the hard-sphere interactions between the solute and the solvent by including the surface free energy and mechanical work of creating a cavity of the solute size in the solvent \cite{Stillinger:1973,Pierotti:1976}.

The SPT model can be used in combination with other solute-solvent nonpolar interactions; e.g.  \cite{Wagoner:2006,Dzubiella:2006,Wei:2009,ZhanChen:2010a}:
\begin{eqnarray} \label{eq1nonpolar}
	G^{\rm{NP}} = \gamma A + p V + \int_{\Omega_s} U d{\bf{r}}, \quad {\bf r} \in {\mathbb R}^3,
\end{eqnarray}
where the first two terms are from SPT and the last term is the free energy due to solvent-solute interactions.
Here, $A$ and $V$ are the surface area and volume of the solute, respectively; $\gamma$ is the surface tension; $p$ is the hydrodynamic pressure; $U$ denotes the solvent-solute non-electrostatic interactions; and $\Omega_s$  is the solvent domain.

In our earlier work, we have shown that the surface area in Eq.~\ref{eq1nonpolar} can be evaluated via a two-dimensional (2D) integral for arbitrarily shaped molecules \cite{Bates:2008,Bates:2009}.
For variation purposes, the total free functional must be set up as a 3D integral in ${\mathbb R}^3$.
To this end, we take advantage of geometric measure theory by considering the mean surface area \cite{Wei:2009} and the coarea formula \cite{coarea}:
\begin{eqnarray} \label{eq4area}
	A=\int_0^1 \int_{S^{-1}(c)\bigcap\Omega} d\sigma dc = \int_{\Omega} |   \nabla S({\bf{r}})   | d{\bf{r}}, \quad {\bf r}\in {\mathbb R}^3,
\end{eqnarray}
where $\Omega$ denotes the whole computational domain and $0 \leq S \leq1$ is a hypersurface or simple surface function that characterizes the solute domain and embeds the 2D surface in ${\mathbb R}^3$; $1-S$ characterizes the solvent domain \cite{ZhanChen:2010a}.
Using the function $S$, the volume in Eq.~\ref{eq1nonpolar} can be defined as:
\begin{eqnarray} \label{eq5vol}
	V = \int_{\Omega_m} d{\bf{r}} = \int_{\Omega}  S({\bf{r}})  d{\bf{r}},
\end{eqnarray}
where $\Omega_m$ is the solute domain.
Note that $\Omega_s \cap \Omega_m$ is not empty because the surface function $S$ is a smooth function, which leads to overlap between $\Omega_s $ and $ \Omega_m$ domains.
The last term in Eq.~\ref{eq1nonpolar} can be written in terms of $S$ as:
\begin{eqnarray} \label{eq6vdw}
	\int_{\Omega_s}  U  d{\bf{r}}=  \int_{\Omega} (1-S({\bf{r}})) U d{\bf{r}}.
\end{eqnarray}

Therefore, we have the following nonpolar solvation free energy functional \cite{Wei:2009,ZhanChen:2010a,ZhanChen:2012}: 
\begin{eqnarray} \label{eqnonpolar}
	G^{\rm NP}&[S] = \int \left\{ \gamma |\nabla S | +   p S +   (1-S)U \right\} d{\bf{r}}
\end{eqnarray}
which is in an appropriate form for variational analysis. 

It is important to understand the nature of the solvent-solute non-electrostatic interaction, $U$.
Assume that the aqueous environment has multiple species labeled by $\alpha$, and their interactions with each solute atom near the interface can be given by: 
\begin{eqnarray} \label{eqnInteractions}
	U &=&\sum_{\alpha}\rho_{\alpha} U_{\alpha} \\ \label{eqnInteractions2}
	&=& \sum_{\alpha}\rho_{\alpha}({\bf r})\sum_{j}U_{\alpha j}({\bf r})
\end{eqnarray}
where $\rho_\alpha({\bf r})$ is the density of $\alpha$th solution component, which may be charged or uncharged, and $U_{\alpha j}$ is an interaction potential between the $j$th atom of the solute and the $\alpha$th component of the solvent.
For water that is free of other species, $\rho_\alpha({\bf r})$ is the water molecule density.
In our earlier work \cite{ZhanChen:2010a,ZhanChen:2010b}, we represented solvent-solute interactions using the Lennard-Jones potential.
The full Lennard-Jones potential is singular and can cause computational difficulties \cite{ZhanChen:2010a}; however, Zhao has proposed a way to improve the integration stability in a realistic setting for proteins \cite{SZhao:2011a}.
However, further mathematical algorithms are needed for this class of problems.
The Weeks-Chandler-Anderson (WCA) decomposition of the potential, which separates the attractive and repulsive components \cite{Weeks:1971}, was also found to provide a good account of the attractive dispersion interaction in our earlier work \cite{ZhanChen:2010a,ZhanChen:2010b}. 

The interaction potential $U$ can be further modified to consider additional interactions, such as 
steric effects \cite{Borukhov:1997} and alternate descriptions of van der Waals interactions.

\begin{figure}[p]
	\begin{center}
		\includegraphics[keepaspectratio,width=0.8\textwidth]{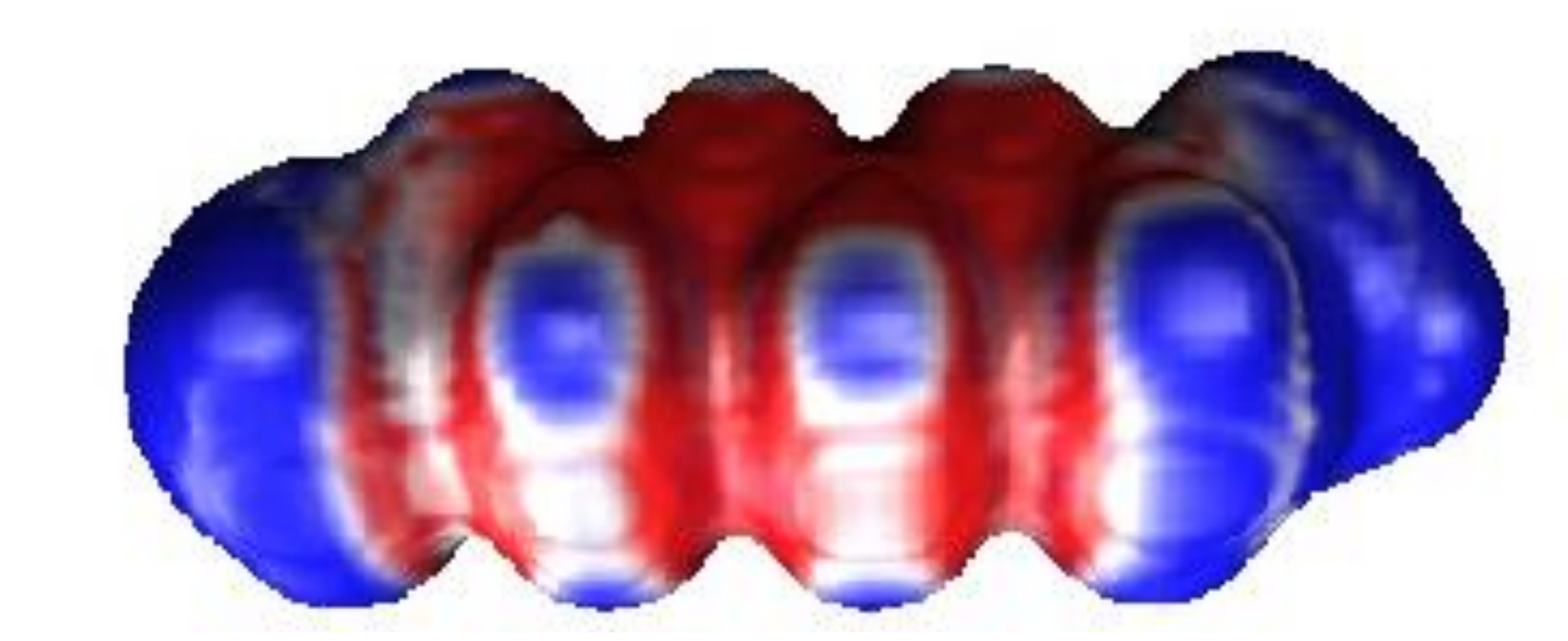}
	\end{center}
	\caption{The final isosurfaces of a nonpolar compound projected with the corresponding van der Waals (vdW) potential for glycerol triacetate \cite{ZhanChen:2012}.}
	\label{figsurface}
\end{figure}

\begin{table}[!p]
	\caption{Solvation energies calculated with the differential geometry nonpolar solvation model for a set of 11 alkanes in comparison with an explicit solvent model \cite{Gallicchio:2000}.
	Errors are computed with respect to experimental data \cite{Cabani:1981}.}
	\begin{center}
		\begin{tabular}{|c|cc|cc|cc|cc|}
			\hline\hline
			&\multicolumn{2}{c|}{Rep. part (kcal/mol)}
			&\multicolumn{2}{c|}{Att. part (kcal/mol)}
			&\multicolumn{2}{c|}{Total (kcal/mol)}
			&\multicolumn{2}{c|}{Error (kcal/mol)}\\
			\hline Compound  &  DG-NP & Explicit  & DG-NP & Explicit
			& DG-NP & Explicit  &  DG-NP & Explicit \\
			\hline
			methane	&4.71	&5.72	&-2.73	&-3.31	&1.98	&2.41	&-0.02	&0.41\\
			ethane	&6.65	&8.07	&-4.75	&-5.44	&1.90	&2.63	&0.07	&0.80\\
			butane	&10.30	&10.10	&-8.18	&-7.21	&2.12	&2.89	&0.04	&0.81\\
			propane	&8.50	&12.19	&-6.45	&-8.98	&2.04	&3.21	&0.08	&1.25\\
			pentane	&12.19	&14.22	&-9.82	&-10.77	&2.37	&3.45	&0.04   &1.12\\
			hexane	&14.03	&16.17	&-11.54	&-12.38	&2.50	&3.78	&0.01	&1.30\\
			isobutane	&10.14	&11.91	&-7.97	&-8.88	&2.16	&3.03	&-0.36	&0.51\\
			2-methylbutane	&11.73	&13.64	&-9.35	&-10.13	&2.38	&3.51	&0.00	&1.13\\
			neopentane	&11.81	&13.62	&-9.20	&-10.39		&2.61	&3.23	&0.11	&0.73\\
			cyclopentane	&10.60	&12.79	&-9.43	&-9.99	&1.17	&2.80	&-0.03	&1.60\\	
			cyclohexane	&12.05	&14.00	&-10.78	&-11.66		&1.27	&2.34	&0.04	&1.11	
			\\ \hline\hline
		\end{tabular}
	\end{center}
	\label{tabledecom}
\end{table}

The Euler-Lagrange equation is used in our variational approach.
By variation of the energy functional with respect to $S$, we arrive at an elliptic equation
\begin{eqnarray}\label{eqElliptic}
	\nabla\cdot\left(\gamma\frac{\nabla S}{|\nabla S |}\right) -p + U = 0,
\end{eqnarray}
where $\nabla\cdot\left(\gamma\frac{\nabla S}{|\nabla S |}\right)$ is a mean curvature term as the surface tension $\gamma$ is treated as a constant.
A standard computational procedure used in our earlier work \cite{Bates:2006,Bates:2008, Bates:2009} involves converting Eq.~\ref{eqElliptic} into a parabolic equation by introducing an artificial time variable:
\begin{eqnarray}\label{eqElliptic1}
	\frac{\partial S}{\partial t}&=&  |\nabla S |\left[ \nabla\cdot\left(\gamma\frac{\nabla S}{|\nabla S |}\right) + V_{\rm NP}\right]
\end{eqnarray}
where $V_{\rm NP}=-p + U$ is a potential-driving term for the time-dependent problem.
Equation \ref{eqElliptic1} is a generalized Laplace-Beltrami equation whose solution leads to the minimization of the nonpolar solvation free energy with respect to the surface function $S$. 
  	
The accuracy of the nonpolar solvation model performance is crucial to the success of other expanded versions of the differential geometry formalism.
In particular, as the electrostatic effect and its associated approximation error are excluded, the major factor impacting the nonpolar solvation model is the solvent-solute boundary, which is governed by the DG-based formalism.
Therefore, the nonpolar model provides the most direct and essential validation of the DG-based models.
In our recent work \cite{ZhanChen:2012}, the DG-based nonpolar solvation (DG-NP) model was tested using a large number of nonpolar compounds. 
Table \ref{tabledecom} presents a small portion of our results \cite{ZhanChen:2012} compared with an explicit nonpolar model \cite{Gallicchio:2000} and experimental data \cite{Cabani:1981}.
The solvation free energy is decomposed into repulsive and attractive parts, showing dramatic cancellations.
The predicted total nonpolar solvation energies are in good agreement with experimental measurements.
More extensive validation of our DG-NP model can be found in an earlier paper \cite{ZhanChen:2012}. 
 
\subsection{Incorporating polar solvation with a Poisson-Boltzmann model}\label{Full}
Most biomolecules are either charged or highly polarized; therefore, electrostatic interactions are indispensable in their theoretical description.
The energy of electrostatic interactions can be modeled by a number of theoretical approaches, including Poisson-Boltzmann (PB) theory \cite{Lamm:2003,Fogolari:2002,Sharp:1990a,Davis:1990a}, polarizable continuum theory \cite{Tomasi:2005,Mei:2006}, and the generalized Born approximation \cite{Dominy:1999,Bashford:2000}.
In our work, we incorporate PB theory for the polar solvation free energy and optimize the electrostatic solvation energy in our variational procedure.

Using the surface function $S$ and electrostatic potential $\Phi$, a PB model for the polar solvation free energy can be expressed by \cite{Wei:2009,ZhanChen:2010a}:
\begin{eqnarray} \label{eq7polar}
	G_{\rm{polar}}=\int \left\{ S\left[  -\frac{\epsilon_m}{2}|\nabla\Phi|^2 + \Phi\ \varrho\right] +
	(1-S)\left[-\frac{\epsilon_s}{2}|\nabla \Phi|^2-k_B T \sum_{\alpha} \rho_{\alpha 0} \left(e^{-\frac{q_{\alpha }\Phi+U_{\alpha} -\mu_{\alpha0}}{k_B T }}-1\right)
	\right]\right\}d{\bf{r}},
\end{eqnarray}
where $\epsilon_s$ and $\epsilon_m$ are the dielectric constants of the solvent and solute, respectively, and $\varrho$ represents the fixed charge density of the solute.
The charge density is often modeled by a point charge approximation $\varrho=\sum_{j}Q_{j} \delta ({\bf{r}}-{\bf{r}}_{j})$, with $Q_{j}$ denoting the partial charge of the $j$th atom in the solute.
$k_B$ is the Boltzmann constant; $T$ is the temperature; $\rho_{\alpha 0}$ denotes the reference bulk concentration of the $\alpha$th solvent species; and $q_{\alpha }$ denotes the charge valence of the $\alpha$th solvent species, which is zero for an uncharged solvent component.
In Eq.~\ref{eq7polar}, the form of the Boltzmann distribution \cite{Wei:2012} is different from that featured in our earlier work \cite{Wei:2009,ZhanChen:2010a}.
\begin{eqnarray}\label{eqn:Boltzman}
	\rho_{\alpha}=\rho_{\alpha 0} e^{-\frac{q_{\alpha }\Phi+U_{\alpha} -\mu_{\alpha0}}{k_B T }}
\end{eqnarray}
 with $\mu_{\alpha0}$ being a relative reference chemical potential that reflects differences in the equilibrium activities of the different chemical species, and thus their concentrations.
The extra term $e^{-\frac{U_{\alpha}}{k_BT}}$ in Eq.~\ref{eqn:Boltzman} describes the solvent-solute interactions near the interface beyond those implicitly represented by $S$.
Therefore, $e^{-\frac{U_{\alpha}}{k_BT}}$ provides a non-electrostatic correction to the charge density near the interface.

The resulting total free energy functional for the full solvation system was first proposed in 2012 \cite{Wei:2012}:
\begin{eqnarray} \label{eq8tot}
	\begin{aligned}
		G^{\rm PB}_{\rm{total}}[S,\Phi]& =  \int\left\{ \gamma |\nabla S | +   p S  
		+S \left[  -\frac{\epsilon_m}{2}|\nabla\Phi|^2 + \Phi\ \varrho\right] \right. \\
		& \left. +(1-S)\left[-\frac{\epsilon_s}{2}|\nabla\Phi|^2-k_B T \sum_{\alpha} \rho_{\alpha 0}\left( e^{-\frac{q_{\alpha }\Phi+U_{\alpha} -\mu_{\alpha0}}{k_B T }}-1\right) \right] \right\} d{\bf{r}}.
	\end{aligned}
\end{eqnarray}
Note that the energy functional (Eq.~\ref{eq8tot}) differs from that in our earlier work \cite{Wei:2009,ZhanChen:2010a} and that of Dzubiella {\it et al.}  \cite{Dzubiella:2006, Dzubiella:2006a} not only in terms of the Boltzmann distribution, but also in the solvent-solute interactions $(1-S)U$, which is omitted in the present form. As shown in Section \ref{sec:multiscale},  the present form is consistent with the DG-based Poisson-Nernst-Planck (PNP) theory at equilibrium.  The DG-based PNP model offers a more detailed description of solvent densities based on fundamental laws of physics. As a result, the formalism  of the DG-based full solvation model should agree with that of the DG-based PNP model at equilibrium.

The total solvation free energy in Eq.~\ref{eq8tot} is expressed as a functional of the surface function $S$ and electrostatic potential $\Phi$.
Therefore, the total solvation free energy functional can be minimized with respect to $S$ and $\Phi$ via the variational principle.
Variation with respect to $S$ leads to:
\begin{eqnarray}  \label{eq9vars}
	\begin{aligned}
		-\nabla\cdot\left(\gamma\frac{\nabla S}{|\nabla S |}\right) + p 
		-\frac{\epsilon_m}{2}|\nabla\Phi|^2 +\Phi\ \varrho 
    	+\frac{\epsilon_s}{2}|\nabla\Phi|^2+ k_B T \sum_{\alpha} \rho_{\alpha 0}\left( e^{-\frac{q_{\alpha }\Phi+U_{\alpha} -\mu_{\alpha0}}{k_B T }}-1\right)=0.
	\end{aligned}
\end{eqnarray}
Using the same procedure discussed earlier, we construct the following generalized Laplace-Beltrami equation:  
\begin{eqnarray}\label{eq10surf}
	\frac{\partial S}{\partial t}&=& |\nabla S |\left[\nabla\cdot\left(\gamma\frac{\nabla S}{|\nabla S |}\right)
   + V_{\rm PB}\right],
\end{eqnarray}
where the potential driven term is given by
\begin{eqnarray}\label{eq11vterm}
	V_{\rm PB}=  -p 
	+\frac{\epsilon_m}{2}|\nabla\Phi|^2 - \Phi\ \varrho-\frac{\epsilon_s}{2}|\nabla\Phi|^2 - k_B T \sum_{\alpha} \rho_{\alpha 0}\left( e^{-\frac{q_{\alpha }\Phi+U_{\alpha} -\mu_{\alpha0}}{k_B T }}-1\right).
\end{eqnarray}
As in the nonpolar case, solving the generalized Laplace-Beltrami equation (\ref{eq10surf}) generates the solvent-solute interface through the function $S$.
Variation with respect to $\Phi$ gives the generalized PB (GPB) equation:
\begin{eqnarray}\label{eq13poisson}
	-\nabla\cdot\left(\epsilon(S) \nabla\Phi\right)= S\varrho
    +(1-S)\sum_{\alpha} q_{\alpha}\rho_{\alpha 0}e^{-\frac{q_{\alpha }\Phi+U_{\alpha} -\mu_{\alpha0}}{k_B T }}, 
\end{eqnarray}
where $\epsilon(S)=(1-S)\epsilon_s+S\epsilon_m$ is the generalized permittivity function.
As shown in our earlier work \cite{Wei:2009,ZhanChen:2010a}, $\epsilon(S)$ is a smooth dielectric function gradually varying from $\epsilon_m$ to $\epsilon_s$.
Thus, the solution procedure of the GPB equation avoids many numerical difficulties of solving elliptic equations with discontinuous coefficients \cite{Zhao:2004,Zhou:2006c,Zhou:2006d,Yu:2007c,Yu:2007a} in the standard PB equation.

Equations \ref{eq10surf} and \ref{eq13poisson} are solved for the surface function $S$ and electrostatic potential $\Phi$, respectively.
These coupled ``Laplace-Beltrami and Poisson-Boltzmann'' equations are the governing equation for the DG-based solvation model in the Eulerian representation.
The Lagrangian representation of the DG-based solvation model has also been derived \cite{ZhanChen:2010b}.
Both the Eulerian and Lagrangian solvation models have been shown \cite{ZhanChen:2010a,ZhanChen:2010b} to be essentially equivalent and provide very good predictions of solvation energies for a diverse range of compounds.

\subsection{Improving Poisson-Boltzmann model charge distributions with quantum mechanics} \label{Quantum}

While our earlier DG-based solvation models resolved the problem of {\it ad hoc} solute-solvent boundaries, they depended on existing force field parameters for atomic partial charge and radius assignments.
Most force field models are parametrized for a certain class of molecules or materials which often limits their transferability and applicability.
In particular, fixed partial charges do not account for charge rearrangement during the solvation process \cite{Ponder:2010,Grossfield:2003,Schnieders:2007}.
Therefore, a quantum solvation model that can self-consistently update the charge density of the solute molecule during solvation offers the promise of improving the accuracy and transferability of our DG-based solvation model.

A quantum mechanical formulation of solute charge density can be pursued in a number of ways.
The most accurate treatment is the one that uses quantum mechanical first principle or {\it  ab initio} approaches.
However, the {\it  ab initio} calculation of the electronic structure of a macromolecule is currently prohibitively expensive due to the large number of degrees of freedom.
A variety of elegant theories and algorithms have been developed in the literature to reduce the dimensionality of this many-body problem \cite{Lee:1996,Yang864575,WeitaoYang:1991,Yang911208,Yang911438,Yang871569,Goedecker:1999,Yang885494}.
In earlier work from the Wei group, a density functional theory (DFT) treatment of solute electron distributions was incorporated into our DG-based solvation model \cite{ZhanChen:2011a}.
In this work, we review the basic formulation and present an improved DG-DFT model for solvation analysis.
Our goal is to construct a DG-DFT based solvation model that will significantly improve the accuracy of existing solvation models and still be orders of magnitude faster than explicit solvent models.

DFT uses functionals of single-electron distributions to represent multi-electron properties so that the total dimensionality is dramatically reduced.
To combine DFT with our DG-based solvation formulation, we define the kinetic energy functional as:
\begin{equation}\label{kinetic}
	G_{\rm kin}[n]=\sum_j\int S({\bf r})\frac{\hbar^2}{2m}|\nabla \psi_j({\bf r})|^2d{\bf r}
\end{equation}
where $n$ is the total electron density, $m({\bf r})$ is the position-dependent electron mass, $\hbar=\frac{h}{2\pi}$ with $h$ being the Planck constant, and $\psi_j({\bf r})$ are the Kohn-Sham orbitals.
The total electron density $n$ is obtained by: 
\begin{equation}\label{density}
	n({\bf r})=\sum_i|\psi_i|^2,
\end{equation}
where the summation is over all of the Kohn-Sham orbitals.

In the absence of external potentials, the electrostatic potential energy of nuclei and electrons can be represented by the Coulombic interactions among the electrons and nuclei.
There are three groups of electrostatic interactions: interactions between nuclei, interactions between electrons and nuclei, and interactions between electrons.
Following the Born-Oppenheimer approximation, we neglect nuclei interactions in our DG-based model.
Using Coulomb's law, the repulsive interaction between electrons can be expressed as the Hartree term:
\begin{equation}\label{ele-ele}
	U_{ee}[n]= \frac{1}{2}\int  \frac{e_C^2n({\bf r})n({\bf r}')}{\epsilon({\bf r})|{\bf r}-{\bf r}'|}d{\bf r}',
\end{equation}
where $e_C$ is the unit charge of an electron; $\epsilon({\bf r})$ is the position-dependent electric permittivity; and ${\bf r}$ and ${\bf r}'$ are positions of two interacting electrons.
Equation~\ref{ele-ele} $U_{ee}[n]$ involves nonlinear functions of the electron density $n$ which implies the need for iterative numerical variational methods, even in the absence of solvent density.
The attractive interactions between electrons and nuclei are given by:
\begin{equation}\label{nucele}
	U_{en}[n]=-\sum_I\frac{e_C^2n({\bf r})Z_I}{\epsilon({\bf r})|{\bf r}-{\bf R}_I |}
\end{equation}
where $Z_I$ is the charge of the nucleus.
The total potential energy functional is then given by
\begin{equation}\label{Elec}
	G_{\rm potential}=\int_\Omega S({\bf r})\left( U_{ee}[n]+U_{ne}[n]+ E_{\rm XC}[n]\right)d{\bf r},
\end{equation}
where the last term, $E_{\rm XC}$, is the exchange-correlation potential, which approximates the many-particle interactions in the solute molecule. 

Intuitively, it appears that the total free energy functional for the DG-based model is the simple summation of the polar, nonpolar, kinetic, and potential energy.
However, such a summation will lead to double counting because of the coupling among different energy terms.
For example, the electrostatic energy depends on the charge density, which, in turn, depends on the kinetic and potential energies of electrons.
Additionally, the electrostatic potential serves as a variable in the polar energy functional and also serves as a known input in the potential energy of electrons through solution of the Poisson equation in vacuum $(\epsilon=1)$
\begin{equation}\label{eqn:Poisson}
	-\nabla^2  \phi_v({\bf r}) = \rho^v_{\rm total}({\bf r}),
\end{equation}
where $\phi_v$ is the electrostatic potential in vacuum and $\rho^v_{\rm total}=n_v+n_n$ with $n_{v}({\bf r})$ being the electron density in vacuum and $n_n$ the density of nuclei.
The solution of the Poisson equation in vacuum is:
\begin{equation}\label{eqn:PotentialV}
	\phi_{v}({\bf r})=\int \frac{e_C n_{v}({\bf r}')}{|{\bf r}-{\bf r}'|}d{\bf r'}-\sum_I\frac{e_C Z_I}{|{\bf r}-{\bf R}_I |}.
\end{equation}
Of note, the solution to Eq.~\ref{eqn:PotentialV} is the exact total Coulombic potential of the electron-electron and electron-nucleus interactions.
Therefore, we do not need to include $U_{ee}[n]$ and $U_{en}[n]$ terms in the total free energy functional.

Based on the preceding discussions, we propose a   total free energy functional for solutes at equilibrium:
\begin{multline}\label{total}
	G_{\rm total}^{\rm DFT-PB}[S,\phi,n] = \int_{\Omega}\left\{\gamma |\nabla S({\bf r}) |  +p S({\bf r}) 
	+ S({\bf r})\left[\rho_{\rm total}\phi-\frac{1}{2}\epsilon_m|\nabla\phi|^{2}\right] \right. \\
	\left. + (1-S({\bf r})) \left[-\frac{\epsilon_s}{2}|\nabla \Phi|^2-k_B T \sum_{\alpha} \rho_{\alpha 0} \left(e^{-\frac{q_{\alpha }\Phi+U_{\alpha} -\mu_{\alpha0}}{k_B T }}-1\right) \right] \right.\\
   + \left. S({\bf r})\left[\sum_j\frac{\hbar^2}{2m}|\nabla \psi_j|^2 +  E_{\rm XC}[n]\right] \right\} d{\bf r},
\end{multline}
where the first row is the nonpolar energy functional; the second row is the electrostatic energy functional; and the last row is the electronic energy functional, which is confined to the solute region by $S({\bf r})$.
As already discussed, the term $\rho_{\rm total}=n_v+n_n$ also contributes to the Coulombic potentials of the electron-electron and electron-nucleus interactions.
This total free energy functional provides a starting point for the derivation of governing equations for the DG-based solvation models, as well as the basis for evaluation of solvation free energies.
 
The governing equations for the DG-based solvation model with quantum mechanical charge distributions are determined by the calculus of variations.
As before, variation of Eq.~\ref{total} with respect to the electrostatic potential $\phi$ gives the generalized Poisson-Boltzmann (GPB) equation \cite{Wei:2009,ZhanChen:2010a}:
\begin{equation}\label{pbDFT}
	-\nabla\cdot(\epsilon(S)\nabla \phi)=S\rho_{\rm total}+(1-S)\sum_{\alpha=1}^{N_c}\rho_{\alpha 0}q_\alpha e^{-\frac{q_{\alpha }\Phi+U_{\alpha} -\mu_{\alpha0}}{k_B T }},
\end{equation}
where the dielectric function is defined as before:  $\epsilon(S)=(1-S)\epsilon_s+S\epsilon_m$.
In a solvent without salt, the GPB equation is simplified to be the Poisson equation:
\begin{equation}\label{LPB}
	-\nabla\cdot(\epsilon(S)\nabla \phi)=S\rho_{\rm total}.
\end{equation}
This equation and Eq.~\ref{pbDFT} are similar to the model described in the previous section (Sec.~\ref{Full}).
However, in the present multiscale model, the charge source $\rho_{\rm total}$ is determined by solving the Kohn-Sham equations rather than by the fixed charges $\rho_{m}=\sum_jQ_j\delta({\bf r}-{\bf r}_j)$.

Variation of Eq.~\ref{total} with respect to the surface function $S$ gives a Laplace-Beltrami equation \cite{Bates:2008,Bates:2009,Wei:2009,ZhanChen:2010a}:
\begin{equation}\label{gfe}
	\frac{\partial S}{\partial t}= |\nabla S |\left[ \nabla\cdot\left(\gamma\frac{\nabla S}{|\nabla S|}\right) + V_{\rm DFT-PB}\right],
\end{equation}
where 
\begin{multline}\label{eqn:V} 
	V_{\rm DFT-PB} =  -p 
	+\frac{1}{2}\epsilon_m|\nabla\phi|^{2}-\frac{1}{2}\epsilon_s|\nabla\Phi|^{2}
	-k_BT\sum_{\alpha=1}^{N_c}\rho_{\alpha0}\left(e^{-\frac{q_{\alpha }\Phi+U_{\alpha} -\mu_{\alpha0}}{k_B T }}-1\right) \\
	- \rho_{\rm total}\Phi
	-\sum_j \frac{\hbar^2}{2m}|\nabla \psi_j|^2 - E_{\rm XC}[n] 
\end{multline}
The electronic potentials in the last row of this equation have relatively small contributions to $V_{\rm DFT-PB}$ at equilibrium due to the fact that they essentially are confined inside the solute molecular domain.  
Note that Eq.~\ref{gfe} has the same structure as the potential-driven geometric flow equation defined in the models presented in earlier in this chapter.
As $t\rightarrow \infty$, the initial profile of $S$ evolves into a steady-state solution, which offers an optimal surface function $S$.

Finally, to derive the equation for the electronic wavefunction, we minimize the energy functional with respect to the wavefunction $\psi_j^*({\bf r})$, subject to the Lagrange multiplier \linebreak ($\sum_i E_i \left(\delta_{ij}- \int S\psi_i({\bf r})\psi_j^*({\bf r})d{\bf r}\right)$) for the orthogonality of wavefunctions to arrive at the Kohn-Sham equation 
\begin{eqnarray} \label{preks}
	\left(-\frac{\hbar^2}{2m}\nabla^2+U_{\rm eff}\right)\psi_j=E_j\psi_j, \quad {\rm with} \quad U_{\rm eff}({\bf r})=q\Phi+V_{\rm XC}[n],
\end{eqnarray}
where the Lagrange multiplier constants $E_i$ can be interpreted as energy expectation values, $V_{\rm XC}[n]=\frac{d E_{\rm XC}[n]}{d n}$, and $q\phi$ is the potential contribution from Coulombic interactions.
These electrostatic interactions can be calculated by the GPB equation (\ref{gfe}) with a given total charge density.
Eq.~\ref{preks} does not directly depend on the solvent characteristic function $S$, so existing DFT packages can be used in our computations with minor modifications.  
 
To integrate our continuum model with standard DFT algorithms, Wei and co-workers introduce the reaction field potential $\Phi_{\rm RF}= \Phi-\Phi_0$ with $\Phi_0$ being the solution of the Poisson equation in homogeneous media \cite{ZhanChen:2011a}.
The reaction field potential is the electric potential induced by the polarized solvent and its incorporation leads to the following effective energy function
\begin{equation}
	U_{\rm eff}({\bf r})=q\Phi+V_{\rm XC}[n]=q\Phi_{\rm RF}+ U^0_{\rm eff}({\bf r})
\end{equation}
where $U^0_{\rm eff}({\bf r})=q\Phi_0 +V_{\rm XC}[n]$ is the traditional Kohn-Sham potential available in most DFT algorithms. 
The reaction field potential also appears in the Hamiltonian of the solute in the quantum calculation \cite{Tannor:1994,MLWang:2006,Bashford:1994} and can be obtained from the electrostatic computation in the framework of the continuum models developed above.
In summary, the inclusion of quantum mechanical charge distributions in the DG-based continuum model involves two components:
1) the classical electrostatic problem of determining the solvent reaction field potential with the quantum mechanically calculated charge density and
2) the quantum mechanical problem of calculating the electron charge density with fixed nucleus charges in the presence of the reaction field potential.
To carry out these computations, an intuitive, self-consistent, iterative procedure can be constructed to solve the quantum equations for the electron distribution and the continuum electrostatic equations for the reaction field potential \cite{Tannor:1994,Gogonea:1999,Tomasi:2005,MLWang:2006,Bashford:1994}.
 
After solving the Kohn-Sham equation,  the QM-based charge density can be incorporated into the solvation model in two different ways.
Our preferred approach is to apply the continuous QM charge density directly to the PB equation as a source term.
However, it is also possible to fit the QM charge density into atomic point charges or multipoles for use as the source term \cite{Geng:2007a, Sigfridsson:1998,HaoHu:2007}.
This second approach is most useful when the DG-DFT scheme is used in conjunction with other molecular simulation approaches, such as MM-PBSA or docking.

\section{Differential geometry-based electrolyte transport models} \label{sec:multiscale}

It is well-known that implicit solvent models use both discrete and continuum representations of molecular systems to reduce the number of degrees of freedom; this philosophy and methodology of implicit solvent models can be extended to more general multiscale formulations.
A variety of DG-based multiscale models have been introduced in an earlier paper of Wei  \cite{Wei:2009}.
Theory for the differential geometry of surfaces provides a natural means to separate the microscopic solute domain from the macroscopic solvent domain so that appropriate physical laws are applied to applicable domains.
This portion of the chapter focuses specifically on the extension of the equilibrium electrostatics models described above to non-equilibrium transport problems which are relevant to a variety of chemical and biological systems, such as molecular motors, ion channels, fuel cells  and nanofluidics, with chemically or biologically relevant behavior that occurs far from equilibrium \cite{Wei:2009, Wei:2012,Wei:2013}.

Another class of DG-based multiscale models involves the dynamics and transport of ion channels, transmembrane transporters and nanofluidics.
In new multiscale models developed by the Wei group, the total energy functionals are modified with additional chemical energies to account for spatially inhomogeneous ion density distribution and charge fluxes due to applied external field gradients and inhomogeneous solvent concentrations across membranes. The Nernst-Planck equation is constructed using Fick's law via a generalized chemical potential governed by the variational principle. Together with the Laplace-Beltrami equation for the surface function and Poisson equation for electrostatic potential, the resulting DG-based PNP theory reduces to our PB theory at equilibrium \cite{Wei:2012}. The PNP equation has been thoroughly studied in the biophysical literature \cite{Im:2002, Gillespie:2003, Eisenberg:1993,Kurnikova:1999,daiguji2004ion,cervera2005poisson,hollerbach2001two,coalson2005poisson}; however, a DG-based formulation of the PNP offers many of the advantages that DG-based solvation models described above provide:  elimination of several \textit{ad hoc} parameters from the model and a framework in which to incorporate more complicated solution phenomena such as strong correlations between ions and confinement-induced ion steric effects. Additionally, compared with conventional PNP models \cite{Im:2002, Gillespie:2003, Eisenberg:1993,Kurnikova:1999,daiguji2004ion,cervera2005poisson,hollerbach2001two,coalson2005poisson}, the DG-based PNP models include nonpolar solvation free energy and thus can be used to predict the full solvation energy against experimental data, in addition to the usual current-voltage curves \cite{Wei:2012}.

\subsection{A differential geometry-based Poisson-Nernst-Planck model}\label{Density}
The GPB and Laplace-Beltrami models discussed in the previous section were obtained from a variational principle applied to equilibrium systems.
For chemical and biological systems far from equilibrium, it is necessary to incorporate additional equations (e.g., the Nernst-Planck equation) to describe the dynamics of charged particles.
Various DG-based Nernst-Planck equations have derived from mass conservation laws in earlier work by Wei and co-workers \cite{Wei:2009, Wei:2012}.
We outline the basic derivation here.
For simplicity in derivation, we assume that the flow stream velocity vanishes ($|{\bf v}|=0$) and we omit the chemical reactions in our present discussion.

The chemical potential contribution to the free energy consists a homogeneous reference term and the entropy of mixing \cite{Fogolari:1997}:
\begin{eqnarray} \label{eq16ent}
	G_{\rm{chem}}=\int  \sum_\alpha \left\{  \left(\mu^0_{\alpha }-\mu_{\alpha 0} \right)  \rho_\alpha + k_B T  \rho_\alpha  {\rm{ln}} \ \frac{\rho_\alpha }{ \rho_{\alpha 0} } - k_B T \left(\rho_\alpha  - \rho_{\alpha 0} \right)   \right\} d{\bf{r}},
\end{eqnarray}
where  $\mu^0_{\alpha}$ is the reference chemical potential of the $\alpha$th species at which the associated ion concentration is $\rho_{0 \alpha }$ in a homogeneous system (e.g., $ \Phi=U_\alpha=\mu_{\alpha 0}=0$).
Here, $k_B  T \rho_\alpha  {\rm{ln}} \ \frac{\rho_\alpha }{ \rho_{\alpha 0} }$ is the entropy of mixing, and  $- k_B T \left(\rho_\alpha  - \rho_{\alpha 0} \right) $ is a relative osmotic term \cite{Manciu:2003}. 
The chemical potential of species $\alpha$ can be obtained by variation with respect to $\rho_\alpha$:
\begin{equation}\label{eqnChemPot}
	\frac{\delta G_{\rm{chem}}}{\delta \rho_\alpha} \Rightarrow
	\mu^{\rm chem}_\alpha= \mu^0_{\alpha }-\mu_{\alpha 0} + k_B T  {\rm{ln}} \ \frac{\rho_\alpha }{ \rho_{\alpha 0}}.
\end{equation}
Note that at equilibrium, $\mu^{\rm chem}_\alpha \neq 0$ and $\rho_\alpha \neq \rho_{\alpha 0}$ because of possible external electrical potentials, charged solutes, solvent-solute interactions, and charged species interactions.
This chemical potential energy term can be combined with the polar and and nonpolar contributions discussed in the previous sections to give a total system free energy of
\begin{multline} \label{eq17tot}
	G^{\rm PNP}_{\rm{total}}[S,\Phi,\{\rho_\alpha\}] = \int \left\{ \gamma |\nabla S | +   p S +   (1-S)U \right. \\
	\left. +S\left[  -\frac{\epsilon_m}{2}|\nabla\Phi|^2 + \Phi\ \varrho\right]
	+(1-S)\left[-\frac{\epsilon_s}{2}|\nabla\Phi|^2 +\Phi\sum_{\alpha} \rho_{\alpha}q_{\alpha} \right]\right.  \\
	\left. +(1-S)\sum_\alpha \left[   \left(\mu^0_{\alpha }-\mu_{\alpha 0} \right) \rho_\alpha + k_B T  \rho_\alpha  {\rm{ln}} \ \frac{\rho_\alpha }{ \rho_{\alpha 0} } - k_B T \left(\rho_\alpha  - \rho_{\alpha 0} \right)  + \lambda_\alpha \rho_\alpha \right] \right\} d{\bf{r}},
\end{multline}
where $\lambda_\alpha$ is a Lagrange multiplier, which is required to ensure appropriate physical properties at equilibrium \cite{Fogolari:1997}.
In this functional, the first row is the nonpolar solvation free energy contribution, the second row is the polar solvation free energy contribution, and the third row is chemical potential energy contribution.
A unique aspect of this PNP formulation is the inclusion of nonpolar solvation free energy contribution to the functional (see Eq.~\ref{eq1nonpolar}).

While electrostatic interactions provide a strong driving force for many biomolecular phenomena, they are not the only source of ion-ion and ion-solute interactions.  
In the heterogeneous environment where biomolecules interact with a range of aqueous ions, counterions, and other solvent molecules, electrostatic interactions often manifest themselves in a variety of different forms related to polarization, hyperpolarization, vibrational and rotational averages, screening effects, etc.
For example, size effects have been shown to play an important role in macromolecular interactions \cite{Hyon:2011,DuanChen:2012b,Bazant:2009,Levin:2002,Grochowski:2008,Vlachy:1999}.
Another important effect is the change of ion-water interactions due to geometric confinement, which is commonly believed to result in channel selectivity for sodium and/or potassium ions \cite{DuanChen:2012b}.
In past papers by Wei and co-workers, these types of interactions are called ``non-electrostatic interactions'' or ``generalized correlations'' \cite{DuanChen:2012b,Wei:2012} and are incorporated into the DG-based models by modifying Eqs.~\ref{eqnInteractions} and \ref{eqnInteractions2}:
\begin{eqnarray} \nonumber
	U&=&\sum_{\alpha}\rho_{\alpha} U_{\alpha}\\ \label{eqnInteractions3}
	U_{\alpha}& =& \sum_{j}U_{\alpha j}({\bf r}) + \sum_{\beta}U_{\alpha \beta}({\bf r}),
\end{eqnarray}
where the subscript $\beta$ runs over all solvent components, including ions and water.
In general, we denote $U_\alpha$ as any possible non-electrostatic interactions in the system.
The inclusion of these non-electrostatic interactions does not change the derivation or the form of other expressions presented in the preceding section.
The total free energy functional (Eq.~\ref{eq17tot}) is a function of the surface function $S$, electrostatic potential $\Phi$, and the ion concentration $\rho_\alpha$. The governing equations for the system are derived using the variational principle.

We first derive the generalized Poisson equation by the variation of the total free energy functional with respect to the electrostatic potential $\Phi$.
The resulting generalized Poisson equation is: 
\begin{eqnarray}\label{eq24poisson}
	-\nabla\cdot\left(\epsilon(S) \nabla\Phi\right)= S\varrho +(1-S)\sum_{\alpha} \rho_{\alpha}q_{\alpha},
\end{eqnarray}
where $\epsilon(S)=(1-S)\epsilon_s+S\epsilon_m$ is an interface-dependent dielectric profile.
The generalized Poisson equation (Eq.~\ref{eq24poisson}) involves the surface function $S$ and the densities of ions $\rho_{\alpha}$, which are to be determined.   
Variation with respect to the ion density $\rho_\alpha$ leads to the relative generalized potential $\mu^{\rm gen}_\alpha$
\begin{equation}\label{eq20varn}
	\frac{\delta G^{\rm PNP}_{\rm{total}}}{\delta \rho_\alpha} \Rightarrow
	\mu^{\rm gen}_\alpha= \mu^0_{\alpha }-\mu_{\alpha 0} + k_B T  {\rm{ln}} \ \frac{\rho_\alpha }{ \rho_{\alpha 0}} + q_{\alpha} \Phi+ U_{\alpha} + \lambda_\alpha
	=\mu^{\rm chem}_\alpha + q_{\alpha} \Phi+ U_{\alpha} +\lambda_\alpha.
\end{equation}
We require $\mu^{\rm gen}_\alpha$, rather than $\mu^{\rm chem}_\alpha$, to vanish at equilibrium.
Therefore, we require: 
\begin{eqnarray}\nonumber
	\lambda_\alpha &=& -\mu^0_{\alpha } \\ \label{eq20Equil}
	\rho_\alpha &=& \rho_{\alpha 0}e^{-\frac{q_{\alpha }\Phi+U_{\alpha} -\mu_{\alpha0}}{k_B T }}.
\end{eqnarray}
Using these relations, the relative generalized chemical potential $\mu^{\rm gen}_\alpha$ can be rewritten as:
\begin{equation}\label{eq21mu}
	\mu^{\rm gen}_\alpha= k_B T {\rm{ln}} \ \frac{\rho_\alpha }{ \rho_{ \alpha 0} }\ + q_{\alpha} \Phi+  U_{\alpha} -\mu_{\alpha0}.
\end{equation}
Wei and co-workers derived a similar quantity from a slightly different perspective in an earlier paper \cite{QZheng:2011b}.  
Note that this chemical potential consists of contributions from the entropy of mixing, electrostatic potential, solvent-solute interaction, and the position-independent reference chemical potential.
For many biomolecular transport problems, diffusion is the major mechanism for transport and relaxation to equilibrium.
By Fick's first law, the diffusive ion flux is ${\bf J}_\alpha=-D_\alpha \rho_\alpha \nabla \frac{\mu^{\rm gen}_\alpha}{k_B T}$ with $D_{\alpha}$ being the diffusion coefficient of species $\alpha$.
The diffusion equation for the mass conservation of species $\alpha$ at the absence of steam velocity is $\frac{\partial \rho_\alpha}{\partial t}=-\nabla \cdot {\bf J}_\alpha$, which results in the generalized Nernst-Planck equation:
\begin{eqnarray}\label{eq22nernst}
	\frac{\partial \rho_\alpha}{\partial t}=\nabla \cdot \left[D_{\alpha} \left(\nabla \rho_{\alpha}+\frac{ \rho_{\alpha}}{k_{B}T}\nabla (q_\alpha\Phi+U_{\alpha})\right)\right],
\end{eqnarray}
where $q_\alpha\Phi+U_{\alpha}$ is a form of the mean field potential.
In the absence of solvent-solute interactions, Eq.~\ref{eq22nernst} reduces to the standard Nernst-Planck equation.
 
Using the Euler-Lagrange equation, one can derive an elliptic equation for the surface function $S$ and, introducing an artificial time as discussed earlier in this chapter, this can be transformed into a parabolic equation:
\begin{eqnarray}\label{eq25surf}
	\frac{\partial S}{\partial t}&=& |\nabla S |\left[\nabla\cdot\left(\gamma\frac{\nabla S}{|\nabla S|}\right) + V_{\rm PNP}\right],
\end{eqnarray}
where the driving term is 
\begin{eqnarray}\label{eq26vterm}
	V_{\rm PNP} &=& -p + U
	+ \frac{\epsilon_m}{2}|\nabla\Phi|^2 - \Phi\ \varrho-\frac{\epsilon_s}{2}|\nabla\Phi|^2+\Phi\sum_{\alpha} \rho_{\alpha}q_{\alpha}  \\ \nonumber
	&& + \sum_\alpha  \left[k_B T  \left( \rho_\alpha  {\rm{ln}} \ \frac{\rho_\alpha }{ \rho_{\alpha 0} } -\rho_\alpha  +\rho_{\alpha 0} \right)  -\mu_{\alpha 0}  \rho_\alpha \right].
\end{eqnarray}
Equations \ref{eq22nernst}, \ref{eq24poisson}, and \ref{eq25surf} form a coupled system of equations describing the surface function $S$, charge concentrations $\rho_\alpha$, and electrostatic potential $\Phi$.
This coupled system differs from the original PNP equations through the coupling of the surface definition are to charge concentrations and electrostatics.
We call this DG-based system the ``Laplace-Beltrami Poisson-Nernst-Planck'' (LB-PNP) model.

In general, the total free energy functional of the DG-based PNP model in Eq.~\ref{eq17tot} differs from that of the DG-based PB model in Eq.~\ref{eq8tot}.
The difference also exists between the surface-driven term $V_{\rm PNP}$ in the charge transport model and $V_{\rm BP}$ in the solvation model.
Moreover, $\rho_{\alpha}$ in the charge transport model is determined by the Nernst-Planck equation (\ref{eq22nernst}) rather than the Boltzmann factor.
However, if the charge flux is zero for the electrodiffusion system, the PNP model is known to be equivalent to the PB model \cite{Roux:2004}.
Note that at equilibrium, the relative generalized potential vanishes everywhere, and the result is the equilibrium constraint given in Eq.~\ref{eq20Equil}.
Therefore, by using the equilibrium constraint, the total free energy functional in Eq. \ref{eq17tot} becomes \cite{Wei:2012}:
\begin{eqnarray} \label{eq30energy}
	G^{\rm PNP}_{\rm{total}} 
	\longrightarrow
	G^{\rm PB}_{\rm{total}}, \quad ~{\rm as} ~\rho_\alpha \longrightarrow \rho_{\alpha 0}e^{-\frac{q_{\alpha }\Phi+U_{\alpha} -\mu_{\alpha0}}{k_B T }}.
\end{eqnarray}
This relationship shows that, under the equilibrium assumption, the total free energy functional for the charge transport model reduces to the equilibrium solvation model presented earlier (Eq.~\ref{eq8tot}).
Furthermore, for the surface-driven functions of the generalized LB equation, it is easy to show \cite{Wei:2012} that under the equilibrium constraint, one has: 
\begin{eqnarray}\label{eq27vterm} 
   V_{\rm PNP} \longrightarrow V_{\rm BP}, \quad ~{\rm as} ~\rho_\alpha \longrightarrow \rho_{\alpha 0}e^{-\frac{q_{\alpha }\Phi+U_{\alpha} -\mu_{\alpha0}}{k_B T }}.
\end{eqnarray}
This consistency between the DG-based PNP and PB models is a crucial aspect of this non-equilibrium theory of charge transport. Numerical simulations in Wei's group have confirmed this consistency \cite{Wei:2012}. 

\subsection{Quantum mechanical charge distributions in the Poisson-Nernst-Planck model}\label{Integrated}

\begin{figure}[p]
	\begin{center}
		\includegraphics[keepaspectratio,width=4.5in]{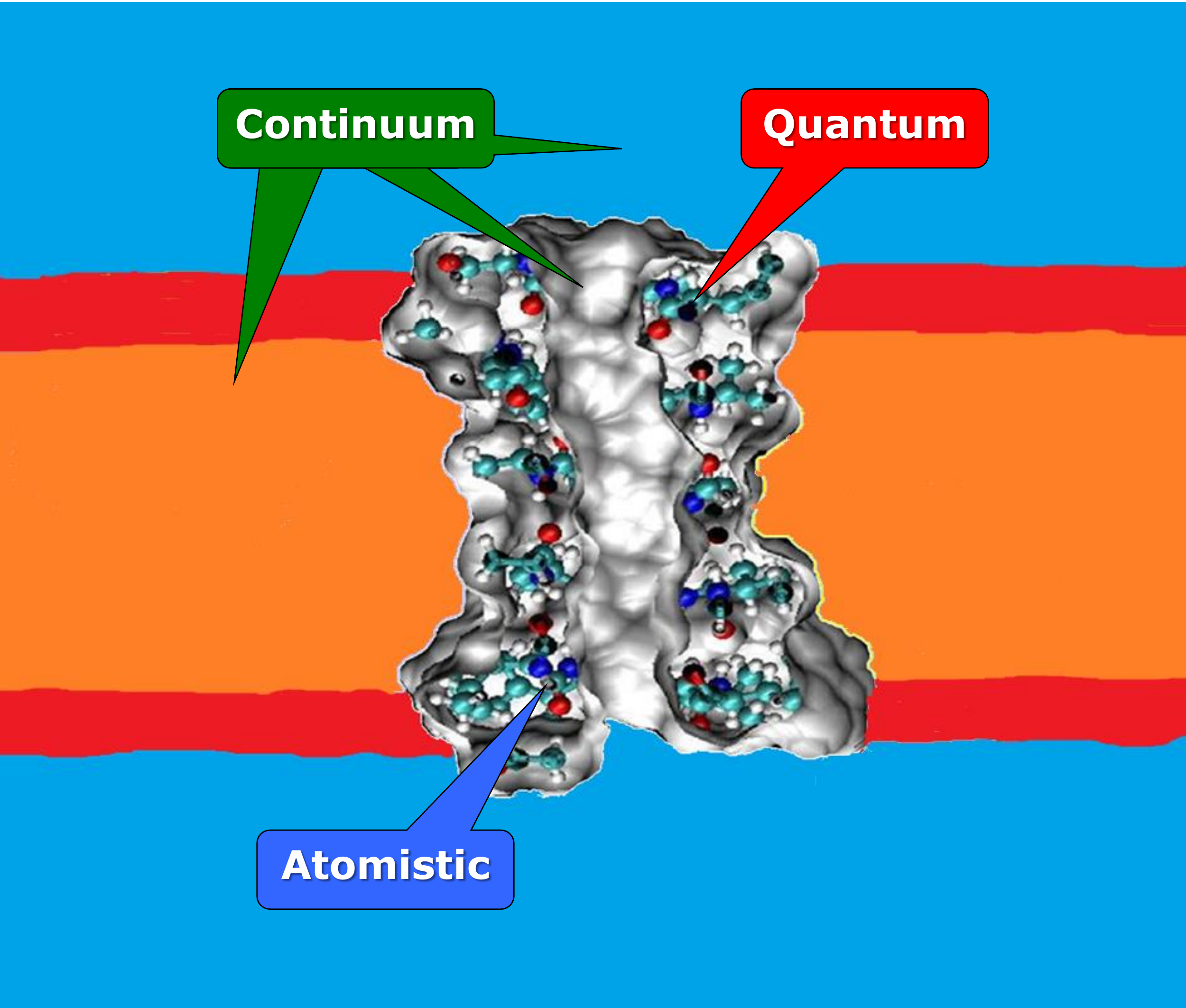}
	\end{center}
	\caption{An illustration of the differential geometry-based DFT-PNP model for ion channels.} \label{figchannel}
\end{figure}

As with the equilibrium solvation models introduced earlier, it is also possible to incorporate quantum mechanical effects into the non-equilibrium transport model.
Our motivation is to account for non-equilibrium ion fluxes and induced response in the electronic structure of the solute or membrane protein.
To this end, we combine our DG-based DFT model with our DG-based PNP model as illustrated in Fig.~\ref{figchannel} to develop a free energy functional and derive the associated governing equations.

The free energy functional is a combination of four models (nonpolar, PB, PNP, and DFT) in a manner which avoids energetic double-counting.
Four variables are used ($S,\Phi,\{\rho_\alpha\}$, and $n$) to minimize the total energy.
The resulting free energy functional has the form:
\begin{multline} \label{eqIntegratetot}
	G^{\rm DFT-PNP}_{\rm{total}}[S,\Phi,\{\rho_\alpha\},n] = \int \left\{ \gamma |\nabla S | +   p S +   (1-S)U \right.  \\
	\left. +S\left[  -\frac{\epsilon_m}{2}|\nabla\Phi|^2 + \rho_{\rm total}\Phi\right] +(1-S)\left[-\frac{\epsilon_s}{2}|\nabla\Phi|^2 +\Phi\sum_{\alpha} \rho_{\alpha}q_{\alpha} \right]\right.  \\
	\left. +(1-S)\sum_\alpha \left[   \left(\mu^0_{\alpha }-\mu_{\alpha 0} \right) \rho_\alpha + k_B T  \rho_\alpha  {\rm{ln}} \ \frac{\rho_\alpha }{ \rho_{\alpha 0} } - k_B T \left(\rho_\alpha  - \rho_{\alpha 0} \right)  + \lambda_\alpha \rho_\alpha \right] \right.\\
	+ \left. S \left[\sum_j\frac{\hbar^2}{2m}|\nabla \psi_j|^2 +  E_{\rm XC}[n]\right] \right\} d{\bf r},
\end{multline}
where the first row is the nonpolar solvation energy functional, the second row is electrostatic energy density of solvation, the third row is the chemical energy functional of solvent ions, and the last row is the energy density of solute electrons in the DFT representation, as explained in earlier sections.
Note that this coupled form places some restrictions on the potential $U$:  in particular, care must be taken to avoid double-counting dispersive and repulsive interactions that are already accounted for in the quantum mechanical treatment.
Using this function, the derivation of governing equations is straightforward.
For the sake of completeness, we discuss all of the governing equations of this new model (as follows).  

As before, variation of the total free energy functional with respect to the electrostatic potential $\Phi$ gives rise to the generalized Poisson equation:
\begin{eqnarray}\label{eqPoissonDFT2}
	-\nabla\cdot\left(\epsilon(S) \nabla\Phi\right)= S \rho_{\rm total} +(1-S)\sum_{\alpha} \rho_{\alpha}q_{\alpha},
\end{eqnarray}
where $\epsilon(S)=(1-S)\epsilon_s+S\epsilon_m$ is an interface-dependent dielectric profile. 
The charge sources in Eq.~\ref{eqPoissonDFT2} are the total charge density $\rho_{\rm total}$ of the solute molecule and the ionic density $\sum_{\alpha} \rho_{\alpha}q_{\alpha}$ of aqueous species.
The former is determined by DFT, while the latter is estimated by the Nernst-Planck theory.
At equilibrium (\ref{eq20Equil}), the generalized Poisson equation (\ref{eqPoissonDFT2}) reduces to the GPB equation given in Eq. \ref{pbDFT}.

The procedure for deriving the Nernst-Planck equation is the same as discussed in the previous section.
We first carry out the variation with respect to $\rho_\alpha$ to obtain the relative generalized potential.
Next, Fick's laws of diffusion are employed to construct the generalized Nernst-Planck equation:
\begin{eqnarray}\label{eq22nernst3}
	\frac{\partial \rho_\alpha}{\partial t}=\nabla \cdot \left[D_{\alpha} \left(\nabla \rho_{\alpha}+\frac{ \rho_{\alpha}}{k_{B}T}\nabla (q_\alpha\Phi+U_{\alpha})\right)\right].
\end{eqnarray}
Formally, this equation has the same form as the generalized Nernst-Planck equation in the last section.
However, to evaluate $U_{\alpha}$, possible effects stemming from the quantum mechanical representation of the electronic structure must be considered. 

As discussed previously, variation with respect to the surface function $S$ leads to a generalized Laplace-Beltrami equation after the introduction of an artificial time:
\begin{eqnarray}\label{eqLB-DFT-PNP}
	\frac{\partial S}{\partial t}&=&|\nabla S |\left[ \nabla\cdot\left(\gamma\frac{\nabla S}{|\nabla S|}\right) + V_{\rm DFT-PNP}\right],
\end{eqnarray}
where the potential driving term is given by 
\begin{multline}\label{eqLB-DFT-PNP2}
	V_{\rm DFT-PNP} = - p + U +\frac{\epsilon_m}{2}|\nabla\Phi|^2 - \Phi\ \varrho-\frac{\epsilon_s}{2}|\nabla\Phi|^2+\Phi\sum_{\alpha} \rho_{\alpha}q_{\alpha}  \\ \nonumber
	+\sum_\alpha  \left[k_B T  \left( \rho_\alpha  {\rm{ln}} \ \frac{\rho_\alpha }{ \rho_{\alpha 0} } -\rho_\alpha  +\rho_{\alpha 0} \right)  -\mu_{\alpha 0}  \rho_\alpha \right] -\sum_j \frac{\hbar^2}{2m}|\nabla \psi_j|^2 - E_{\rm XC}[n].
\end{multline}
At equilibrium (Eq.~\ref{eq20Equil}) $V_{\rm DFT-PNP}$ becomes $V_{\rm DFT-PB}$.
Eq.~\ref{eqLB-DFT-PNP} is coupled to all other quantities, $\Phi, \rho_\alpha$ and $n$.
Fast solutions to this type of equation remains an active research issue \cite{Bates:2009, ZhanChen:2010a, WFTian:2014}.

In the present multiscale DFT formalism, the governing Kohn-Sham equation is obtained via the minimization of the energy functional with respect to $\psi_j^*({\bf r})$, subject to the Lagrange multiplier ($\sum_iE_i \left(\delta_{ij}- \int S\psi_i({\bf r})\psi_j^*({\bf r})d{\bf r}\right)$),  
\begin{eqnarray} \label{preks2}
	\left(-\frac{\hbar^2}{2m}\nabla^2+q\Phi+V_{\rm XC}[n]\right)\psi_j=E_j\psi_j. 
\end{eqnarray}
Although the Kohn-Sham equation does not explicitly involve the surface function and ion densities, the electrostatic potential energy $q\Phi$ is calculated by the GPB equation (\ref{eqPoissonDFT2}) which is coupled with solvent charge density and surface function.
As such, electronic response to ion fluxes in the ion channel is included in the present model.

Equations \ref{eqPoissonDFT2}, \ref{eq22nernst3}, \ref{eqLB-DFT-PNP}, and \ref{preks2} form a complete set of governing equations which are strongly coupled to each other.
Therefore, these equations can be solved by nonlinear iterative procedures \cite{DuanChen:2013,DuanChen:2012a,DuanChen:2012b}  and efficient second-order algorithms \cite{ZhanChen:2010a,ZhanChen:2010b,ZhanChen:2011a,ZhanChen:2012}.
 
\section{Concluding remarks}\label{Concluding}

Geometric analysis, which combines differential geometry (DG) with partial differential equations (PDEs), has generated great successes in the physical sciences and engineering.
In the past decade, DG-based solvation models have been introduced for biomolecular modeling.
This new methodology has been tested over hundreds of molecular test cases, ranging from nonpolar molecules to large proteins.
Our DG-based solvation models use the differential geometry of surfaces theory as a natural means to separate microscopic domains for biomolecules from macroscopic domains for solvents and to couple continuum descriptions with discrete atomistic or quantum representations.
The goal of our  DG-based formalism is to achieve an accurate prediction of essential physical observables while efficiently reducing the dimensionality of complex biomolecular systems.
An important technique used in our  approach is the construction of total free energy functionals for various biomolecular systems, which enables us to put various scales into an equal footing.
Variational principles are applied to the total energy functional to derive coupled governing PDEs for biomolecular systems.

This chapter has focused on equilibrium and non-equilibrium models of electrolyte solutions around biomolecules.
However, the Wei group has also extended this formalism to the multiscale modeling of other systems and biological processes.
One class of multiscale models developed in the Wei group is a DG-based quantum treatment of proton transport \cite{DuanChen:2012a,DuanChen:2012b}.
Proton transport underpins the molecular mechanisms in a variety of systems, including transmembrane ATPases as well as other proton pumps and translocators \cite{HNChen:2007}.
The significant quantum effects in proton permeation require quantum mechanical models, while the large number of degrees of freedom demands a multiscale treatment \cite{Nagle:1978,Roux:1996}.
In the multiscale approach developed by the Wei group, a new DFT is formulated based on Boltzmann statistics, rather than Fermi-Dirac statistics, for protons in the solvent while treating water molecules as a dielectric continuum.
The membrane protein is described in atomistic detail and densities of other ions in the solvent are approximated via Boltzmann distributions, following an approach introduced in our earlier Poisson-Boltzmann-Nernst-Planck theory \cite{QZheng:2011b}.
The resulting multiscale proton model provides excellent predictions of experimental current-voltage relationships  \cite{DuanChen:2012a,DuanChen:2012b}.
Another class of DG-based multiscale models has been proposed by Wei {\it et al.}~for alternative MM and/or continuum elasticity (CE) description of solute molecules, as well as continuum fluid mechanics formulation of the solvent \cite{Wei:2009,Wei:2012,Wei:2013,KLXia:2013d}.
The idea is to endow the DG-based multiscale paradigm with the ability to handle excessively large macromolecules by elasticity description, manage conformational changes with MM, and deal with macromolecular-flow interaction via fluid mechanics.
The theory of continuum elasticity with atomic rigidity (CEWAR) also has been introduced \cite{KLXia:2013d} and treats the molecular shear modulus as a continuous function of atomic rigidity.
Thus, the dynamic complexity of integrating time-dependent governing equations for a macromolecular system is separated from the static complexity of determining the flexibility at given time step. 
In CEWAR, the more time-consuming dynamics is approximated using continuum elasticity theory while the less-time-consuming static analysis is pursued with atomic description.
A recent multidomain formulation by Wei and co-workers allows each different part of a macromolecule to have a different physical description \cite{Wei:2013}. 
Efficient geometric modeling strategies associated with DG-based multiscale models have been developed in both Lagrangian-Eulerian \cite{XFeng:2012a, XFeng:2013b} and Eulerian representations \cite{KLXia:2014a}.
Algorithms for curvature evaluation and volumetric and surface meshing have been developed for organelles, subcellular structures, and multiprotein complexes \cite{XFeng:2012a} and have been combined with electrostatic analysis for the prediction of protein-ligand binding sites \cite{KLXia:2014a}.

\section*{Acknowledgments}
This work was supported in part by National Science Foundation grants IIS-1302285 and DMS-1160352, as well as National Institutes of Health Grant R01GM-090208.
The authors are indebted to their collaborators who have contributed to the DG-based biomolecular modeling.
 
\clearpage
\section*{Literature cited}
\renewcommand\refname{}

\bibliographystyle{unsrt}
\bibliography{refs}

\end{document}